\documentstyle[epsfig,emulateapj]{article}
\def\ltsima{$\; \buildrel < \over \sim \;$}
\def\lsim{\lower.5ex\hbox{\ltsima}}
\def\gtsima{$\; \buildrel > \over \sim \;$}
\def\gsim{\lower.5ex\hbox{\gtsima}}
\def\bi#1{\bbox{#1}}

\def\om{\Omega_m}
\def\hth{\hat{\phi}}
\newcommand{\bm}[1]{\mbox{{\it \boldmath$#1$}}}
\begin{document}
\title{Magnification effects as measures of large-scale structure}
\author{Bhuvnesh Jain} 
\affil{Dept. of Physics and Astronomy, University of Pennsylvania, 
Philadelphia, PA 19104\\
Email: bjain@physics.upenn.edu}

\def\bi#1{\hbox{\boldmath{$#1$}}}
\begin{abstract}

The magnification effects of clustered matter produce variations
in the image sizes and number density of galaxies across the sky.  
This paper advocates the use of these effects in 
wide field surveys to map large-scale structure and the profiles 
of galaxy and cluster sized halos. 
The magnitude of the size variation as a function of angular scale is 
computed and the signal-to-noise is estimated for different survey parameters. 
Forthcoming surveys, especially well designed space-based imaging surveys, 
will have high signal-to-noise on scales of about 0.1 arcminute to 
several degrees. Thus the clustering of matter could be measured on 
spatial scales of about 50 Kpc to 100 Mpc. The signal-to-noise is dominated 
by sample variance rather than shot-noise due to the finite number 
density of galaxies, hence the accuracy of the measurements
will be limited primarily by survey area, sampling
strategy and possible systematics. Methods based on magnification are 
compared with the use of shape distortions and the contrasts and
complementarities are discussed. Future work 
needed to plan survey strategy and interpret measurements 
based on magnification is outlined. 

\end{abstract}


\section{Introduction}

Gravitational lensing refers to the distortions in images of
distant galaxies due to the deflection of light rays by mass
concentrations. The distortion on a circular image can be decomposed 
into an amplification of the size of the image and an anisotropic 
stretching of its shape into an ellipse. The size amplification is called
magnification and the anisotropic stretching is the shear. 
Gravitational lensing due to galaxy clusters and large-scale structure
typically leads to distortions of order 1-10\% (e.g. Gunn 1967; 
Miralda-Escude 1991; Blandford et al 1991; Kaiser 1992; Bernardeau, 
van Waerbeke \& Mellier 1997; Jain \& Seljak 1997; Kaiser 1998). 
In this regime of weak
lensing the magnification $\mu$ is given by
\begin{equation}
\mu = \left((1-\kappa)^2 - |\bm{\gamma}|^2\right)^{-1} \simeq 1 + 2 \kappa . 
\end{equation}
where $\kappa$ is the convergence and $\bm{\gamma}$ the complex shear. So far
observational studies of weak lensing have largely used the measured
ellipticities to estimate the shear and thus the projected mass 
distribution. %This has been successfully done for several clusters 
%and recently in blank fields as well. Thus weak lensing has emerged as
%a promising way to measure dark matter properties without the use of
%biased tracers such as galaxies associated with the dark matter. 

This paper makes the case for using effects of magnification in
addition to the shear in mapping dark matter. Magnification leads to
fluctuations in the sizes and number densities (in a flux limited
survey) of galaxies (e.g. Bartelmann \& Narayan 1995; Broadhurst, 
Taylor \& Peacock 1995; Schneider, King \& Erben 200). 
In the context of galaxy clusters the change in number
density has been used to constrain the mass distribution, but with
less accuracy than the shape measurements (Taylor et al 1998). 
We argue that for forthcoming 
blank field surveys the prospects are much better 
than for clusters to measure both effects of magnification, on sizes 
and number densities. 

(i) Magnification effects, unlike the shear, require control fields to
estimate the mean, unlensed size distribution. This had been a limitation
for small, arcminute sized, cluster fields, but is automatically
done in a blank field survey. 

(ii) The signal-to-noise (henceforth ${\cal S/N}$) 
due to shot noise is somewhat lower for magnification effects
than the shear, and this has proven critical in cluster lensing, since
a factor of 2 in ${\cal S/N}$ is hard to make up. However field
lensing surveys with areas larger than 10 square degrees are mostly in
the regime where sample variance or systematics 
dominate the errors. It is therefore feasible that from forthcoming 
imaging surveys with good control
of systematics (photometric calibration for number density, resolution
for sizes, and psf anisotropy for shear) 
all three lensing measurements can be made. 
Consistency checks on the different systematics can then be made, 
the ${\cal S/N}$ on the measured dark matter clustering improved, 
and new information on halo properties can be extracted. 

(iv) Space based imaging surveys will make possible the measurements
of sizes with an accuracy hard to achieve from the ground. Such
surveys will become feasible over small areas with the Advanced 
Camera for Surveys on the HST, and over substantial fractions of
the sky with a wide field imaging satellite telescope. 

The main goal of this paper is to propose that 
measurements of magnification effects, in particular the effect on 
galaxy sizes, be an integral part of the lensing agenda for 
forthcoming imaging surveys: wide area, multi-color ground based 
surveys like the CFHT Legacy Survey 
(see www.cfht.hawaii.edu/Science/CFHLS/), the 
proposed LSST (www.lss.org) and  
WFHRI (www.ifa.hawaii.edu/ $\tilde{}$kaiser/wfhri) surveys, and 
especially a space based imaging survey as proposed
for the SNAP satellite (http://snap.lbl.gov) which will have the key
requirements of small psf and pixels $\sim 0.1$ arcsecond, photometric
redshifts, and survey area exceeding 100 square degrees (G. 
Bernstein, private communication). 
The formalism for computing statistical measures of magnification is
presented in Section 2. 
%The predicted rms from popular
%cosmological models is shown as a function of angular scale. 
Section 3 provides 
estimates of the ${\cal S/N}$ expected for measurements of size fluctuations. 
We conclude in Section 4. 

\section{Statistical Measures of Magnification}

\subsection{Fluctuations in the size distribution}

The lensing effect of an overdensity in the mass distribution is to
increase the area of a given patch on the sky. The size of a given 
galaxy therefore increases. 
%In the weak lensing regime, the
%magnification $\mu$ can be expressed in terms of the convergence
%$\kappa$ as 
%\begin{equation}
%\mu = \left[(1-\kappa)^2 - |\gamma|^2\right]^{-1} \simeq 1 + 2 \kappa . 
%\end{equation}
The area $A$ and characteristic radius $R\propto A^{1/2}$ 
of a galaxy, for $\kappa \ll 1$,  is then given by:
\begin{equation}
A \to A (1+2\kappa); \ \ R \to R (1+\kappa) .
\label{sizemag}
\end{equation}
Thus the logarithm of the sizes is shifted linearly as
$ {\rm log} \ R \to {\rm log} \ R + \kappa. $

Following the notation of Jain \& Seljak (1997) we introduce
the unperturbed metric
%\begin{equation}
$
ds^2=a^2(\tau)\left[ -d\tau^2+ d\chi^2+r^2(d\theta^2+\sin^2 \theta d\phi^2)
 \right]  ,
%\end{equation}
$
with $\tau$ being conformal time, 
$a$ the expansion factor normalized to unity today, $\chi$ 
the radial comoving distance and $r(\chi)$ the comoving 
angular diameter distance. 
%\begin{eqnarray}
%r(\chi)=\sin_K\chi \equiv
%\left\{ \begin{array}{ll} K^{-1/2}\sin K^{1/2}\chi,\ K>0\\
%\chi, \ K=0\\
%(-K)^{-1/2}\sinh (-K)^{1/2}\chi,\ K<0 
%\label{rchi}
%\end{array}
%\right.
%\end{eqnarray}
%where $K$ is the spatial curvature given by 
%$K=-H_0^2(1-\Omega_m-\Omega-\Lambda)$ with $H_0$ being the 
%Hubble parameter today. 
The convergence $\kappa$ is a weighted projection of the mass density
along the line of sight. It can be expressed as
\begin{equation} 
\kappa_i(\hth)=\frac{3}{2} \om \int_0^{\chi_H} d \chi\ g_i(\chi)
\frac{\delta(r \hth,\chi)}{a}  , 
\label{kappa1}
\end{equation}
where $\chi_H$ denotes the distance to the horizon. 
With $W(\chi)$ denoting the radial distribution 
of galaxies in the sample, the radial weight function $g(\chi)$ 
is given by
\begin{equation}
g(\chi)= r(\chi) \int_\chi^{\chi_H}
{r(\chi' -\chi) \over r(\chi')}W(\chi')d\chi'\ .
\end{equation}
%For a delta-function distribution of galaxies,
%$W(\chi')=\delta_D(\chi'-\chi_S)$, and $g(\chi)$ reduces to 
%$g(\chi)=r(\chi)r(\chi_S-\chi)/r(\chi_S)$. 

%where the comoving radial coordinate $\chi$ and the comoving angular
%diameter distance $r(\chi)$ are explicitly defined in the appendix. 
%$a$ is the expansion factor normalized to unity today and $\chi_H$ is 
%the distance to the horizon.
%The radial weight function $g(\chi)$ can be expressed in terms
%of $r(\chi)$; for a delta-function distribution of source galaxies at
%$\chi_S$ it is $g(\chi)=r(\chi)r(\chi_S-\chi)/r(\chi_S)$.

The variance in the size fluctuations can be related to the variance
in the smoothed convergence by considering the mean size $\bar R_\theta$ 
in a circular aperture of angle $\theta$. If the unlensed or intrinsic 
mean size in such an aperture is denoted $\bar R^i_\theta$, then 
we can define the size shift $\delta R \equiv (\bar R_\theta -  
\bar R^i_\theta)/\bar R^i_\theta$. 
Using equations \ref{sizemag} and \ref{kappa1} the variance of the 
shift in sizes along different lines of sight can be obtained in terms 
of the power spectrum of density fluctuations as
\begin{eqnarray}
\langle{\delta R}^2\rangle(\theta) 
&=&36 \pi^2 \ \Omega_m^2
\int_0^{\infty}\ k dk\int_0^{\chi_0}\ a^{-2}(\chi)\ \nonumber \\
&\times &  P_{\delta}(k,\chi)
\ g^2(\chi)
U^2[kr(\chi)\theta]d\chi,
\label{sizevar}
\end{eqnarray}
where $U(x)=2J_1(x)/x$, with $J_1(x)$ being the Bessel function of
first order. 

\subsection{Cross-correlations induced by magnification}

Magnification causes the observed area of a given
patch of sky to increase, tending 
to dilute the number density, but galaxies fainter than
the limiting magnitude are brightened and may be
included in the sample, thus increasing the number density. The net
effect, known as magnification bias, can go either way depending on 
the slope $s$ of the number counts of galaxies
$N_0(m)$ in a sample with limiting magnitude $m$, $s= d\log N_0/dm$.
Magnification by amount $\mu$ changes the number counts to
(e.g. Broadhurst, Taylor \& Peacock 1995),
$
N^\prime(m)=N_0(m) \mu^{2.5s-1}.
$
In the weak lensing regime, this reduces to
$
N^\prime(m)=N_0(m)\left[ 1+5(s-0.4)\kappa \right].
$
Variations in the number density which are correlated over some
angular separation are produced due to the spatial correlations of the
lensing dark matter. These correlations are difficult to detect since
the galaxies have a strong auto-correlation function due to their
spatial clustering. However the cross-correlation of two
galaxy samples with non-overlapping redshift distributions isolates
the effect of magnification bias. 

The cross-correlation of a foreground-background galaxy sample 
can be obtained in the Limber approximation (Moessner \& Jain 1998): 
\begin{eqnarray}
\omega_{\times}(\theta)&=& 3 \om (2.5 s_2-1) 4 \pi^2 \int_0^{\chi_H} d \chi
W_1(\chi) \frac{g_2(\chi)}{a} \nonumber \\
&& \times \int_0^\infty dk\, k\, P_{\times}(\chi, k)\,
J_0\left[k r(\chi)\theta\right] \; ,
\label{omegagl}
\end{eqnarray}
where the subscripts $1,2$ denote the foreground and background populations
respectively and $P_{\times}(\chi, k)$ is the projected
galaxy-mass cross-power spectrum. 
$\omega_{\times}(\theta)$ can also be measured by 
the quasar-galaxy correlations that have been extensively discussed in 
the literature. The sizes of background galaxies discussed in the
previous sub-section can be used as well; this would alter the equation
above only in the numerical coefficient on the right-hand side. 

Note that $\omega_{\times}(\theta)$ is a measure of
the galaxy-mass cross-correlation. It is the counterpart
of galaxy-galaxy lensing, with the difference that the convergence is
measured rather than the tangential shear. 
Hence it provides a more local
measure of the galaxy-mass cross-correlation, which in the small scale
regime probes the structure of galactic halos. For galaxy clusters one
can measure size increases or number counts of background galaxies 
around individually clusters from high quality data, else they can
be stacked like the galaxies. Large catalogs of clusters will soon
be available for such measurements, and conversely, mass selected
cluster catalogs may be obtainable from these measures of the convergence. 

\subsection{Corrections of high redshift supernovae magnitudes}

Recently Dalal et al (2002) have estimated the capability of shear maps 
to correct the lensing induced dispersion in the measured magnitudes
of high redshift supernovae. The idea is the following. Magnification
effects contribute to the measured scatter in the magnitudes of high
redshift supernovae. Since the lensing contribution can equal or exceed
the intrinsic dispersion for supernovae at $z \gsim 1$, it is valuable
to be able to measure the lensing effect along the lines of sight to
individual supernovae from another tracer and thus correct the supernovae
magnitudes. Dalal et al (2002) needed to assume that the convergence
can be reconstructed on arcminute scales from ellipticity data, which 
may not prove to be feasible as they discuss. Measurements of size 
fluctuations however could directly map the convergence on arcminute 
scales around the line of sight to supernovae, allowing for a
reduction in the scatter of supernova magnitudes. 

The main open questions are: How large is the
shot noise effect for a given survey? How strongly is the 
smoothed convergence estimated from
source galaxies correlated with the value along the line of sight to 
individual supernovae? On arcminute scales ray
tracing simulations can be used to estimate how much stronger 
this correlation is than assumed in 
the Gaussian limit taken by Dalal et al (2002). Hence a quantitative
study is merited to check if one can do better than the reduction of
about 10\% in the lensing dispersion reported by these authors. 

\section{Signal-to-noise estimates}

We use as the primary observable the log of the linear size (such as 
half-light radius). To measure a lensing signal, it is best to select
galaxies on surface brightness which is conserved by lensing
(Bartelmann \& Narayan 1995). In the
following we will assume that photometric redshifts are measured for
the galaxies and that the surface brightness is not contaminated by
atmospheric or instrumental effects. The number density of usable 
galaxies will depend on how conservative the catalog selection for
a specific instrument and survey will need to be. 
The ${\cal S/N}$ in the variance on smoothing
scale $\theta$ using galaxy sizes as an 
estimator of the magnification is then given in terms of the standard 
deviation $\sigma_i$ in the intrinsic solid angles and the number of galaxies
$N_\theta$ per circle of size $\theta$. For a single field of size 
$\theta$ the ${\cal S/N}$ is: 
\begin{equation}
\left( {\cal S \over N}\right)_{1-\rm field} = {\sigma^2_\kappa(\theta) \over \sigma_i^2(I)/N_\theta} \ ,
\label{sn_size}
\end{equation}
where $\sigma^2_\kappa(\theta) = \langle \kappa^2 \rangle$ is the 
variance in the convergence with a top-hat smoothing of angular size 
$\theta$, and $\sigma_i(I)$ is the standard deviation of the size distribution
in a given bin in physical surface brightness. This differs from the
corresponding expression for ellipticity measurements because the
denominator contains a different
$\sigma^2_\epsilon/N^\epsilon_\theta$ 
(the signal due to the variance of $\kappa$ or $\gamma$ in the numerator
is equal in the weak lensing regime). From ground based data, the ${\cal S/N}$ 
for the ellipticity is larger than for the size since psf smearing 
directly affects the size estimate and affects the shape only at second order. 
From space based data however it 
is hard to say a priori whether the ${\cal S/N}$ from shape measures 
is higher (at least by larger than a factor of two) for a given survey area. 
In any case, as shown
below, on large scales sample variance dominates the statistical errors. 

For cosmological measurements, the data size of interest is a wide field
survey from which the variance of $\kappa$ can be measured over a range of
angular scales. Thus there are two angles, the first denoted $\Theta_0$
gives the size of the survey, and the second is the angular scale
 on which the variance is measured, which we will continue to denote
$\theta$. Let $N_f = \Theta_0^2/\theta^2$ be the number of patches of
size $\theta$ used in the measurement of $\kappa^2$. Thus the total number
of galaxies is $N_t = N_f N_\theta$. In the following we will assume
that the $N_f$ patches are uncorrelated. The contribution to the measured
variance due to sample variance and the intrinsic scatter in the size 
distribution is: 
%\begin{equation}
$ {\cal N} = \left[\sqrt{2}\ \sigma^2_\kappa(\theta) +\sigma_i^2/N_\theta\right] / 
\sqrt{N_f}, 
$
%\label{noise_fields}
%\end{equation}
where the first term is the sample variance contribution assuming
a Gaussian distribution, while the second is the contribution from
the intrinsic scatter in the size distribution. 

The ${\cal S/N}$ for the measured variance is then given by 
\begin{equation}
{\cal S \over N} = \sigma^2_\kappa(\theta) \ { \sqrt{N_f} \over 
\left[\sqrt{2}\ \sigma^2_\kappa(\theta) +\sigma_i^2/N_\theta\right] }\ .
\label{sn_fields}
\end{equation}
The above estimate ignores the effect of the kurtosis on the 
sample variance and thus underestimates the sample variance 
on small scales. The effect can be estimated by using the
results of Takada \& Jain (2002) who find that the kurtosis
parameter defined as $S_4 = \langle \kappa^4 \rangle / \sigma^6_\kappa
= 3\times 10^4$ between $1' < \theta < 10'$ and falls off on larger
scales (see their Figure 9). 
In the sample variance contribution, the relevant ratio
is the standard definition of kurtosis in statistics, 
$ \langle \kappa^4 \rangle / \sigma^4_\kappa$, so we need
to find the angular scales on which this ratio is of 
order unity. Over the scales of interest, 
$\sigma^2_\kappa \simeq 3 \times 10^{-4}
\ \theta^{-1}$, where $\theta$ is in arcminutes (Jain \& Seljak 1997). 
Hence we obtain $ \langle \kappa^4 \rangle /  \sigma^4_\kappa \simeq 
10/\theta(')$, a simple expression that is sufficiently
accurate for our purpose. Thus for $\theta \lsim 10'$, the kurtosis
term is important and could increase 
the sample variance by up to a factor of two. 
However, we will see below that 
the shot noise term dominates the sample variance term on scales
smaller than a few arcminutes. 
Hence it is only over a small range in angle, and at worst by a factor
of two, that we have underestimated the sample variance.  
Note that analogous expressions to equation \ref{sn_fields} 
hold for the ${\cal S/N}$ from shape
measurements (e.g. Jain \& Seljak 1997; Schneider et al 1998; 
Hu \& Tegmark 1999). 

It is interesting to consider the relative contributions of 
sample variance and intrinsic scatter to the noise term in
equation \ref{sn_fields}. Again using the approximate 
relation $\sigma_\kappa^2 \propto 1/\theta$, we see 
that the shot-noise term scales $1/\theta^3$, while the sample
variance scales as $1/\theta^2$. Thus on small scales the shot-noise term 
dominates, while on scales larger than a few arcminutes (depending
on the number density of galaxies) the sample variance term dominates. 

Figure 1 shows the predicted variance in size shifts  
and the ${\cal S/N}$ expected for
different survey parameters. We assume a flat 
$\Lambda$CDM model with $\sigma_8=0.9$ and assume 
that photometric redshifts are available for a source 
redshift distribution of the form
$n(z)\propto z^2 {\rm exp}\left[-(z/z_0)^{1.2}\right]$. Varying $z_0$ changes
the mean redshift of the distribution. 
The left panel shows the variance in the
size shift and the two sources of noise: the intrinsic dispersion
of galaxy sizes and sample variance. On scales of order $1'$ and
smaller, the intrinsic dispersion dominates, while on larger scales
sample variance is the main source is noise. In the right panel
the ${\cal S/N}$ achievable with a filled survey is shown. The 
middle solid curve assumes a number density of galaxies of 40 per
square arcminute, a total area of 100 square degrees and an intrinsic
dispersion $\sigma_i=0.5$ (Narayan \& Bartelmann 1995). 
This curve shows that high ${\cal S/N}$ measurements
can be made on scales of order $0.1'-100'$, which corresponds to spatial
scales of about 50 Kpc to 50 Mpc. If the level
of systematics is not a show-stopper, then one can extend
the measurements to larger scales by sparse sampling. For given 
survey area, sparse sampling would increase the $N_f$ term on large
scales. Kaiser (1998) uses a power spectrum analysis to examine
the best strategy  for sparse sampling. 

The lower and upper solid curves in the right panel of Figure 1 
show the effect of changing the survey area by a
factor of ten --- the curves shift up and down by the square root
of the area. The dashed and dot-dashed curves show the effect on
the ${\cal S/N}$ of lower galaxy number density and higher mean
redshift of source galaxies, respectively. If the effective 
number density for which 
sizes can be measured is decreased to 20 per square arcminute, then
the shot noise term on small scales ($\theta < 5'$)
lowers the ${\cal S/N}$. For a higher redshift distribution of
source galaxies, keeping other parameters constant, the signal is
higher, so the ${\cal S/N}$ improves on small scales 
as shown by the dot-dashed curve. If neighboring 
fields are correlated, then the sample variance estimate must be
revised because the effective number of independent fields of given
angular size $\theta$ is smaller. As discussed above, this would lower
the ${\cal S/N}$ for $\theta < 10'$. 
It is clear from the range of the effective parameters in the 
${\cal S/N}$ explored here that even in a conservative scenario,
a survey with area of order 100 square degrees will provide high
${\cal S/N}$ measurements over several decades in length-scale. 

\section{Discussion}

What kind of survey would be suitable 
for measuring the magnification effects discussed in this paper? 
For the effect of magnification on galaxy sizes, a wide area space
based multi-color imaging survey would be ideal. 
It is challenging for a ground 
based telescope to overcome the effect of psf smearing on the size
distribution, unless one has the luxury of a large enough sample of 
galaxies with sizes larger than the psf. With appropriate multi-color 
imaging one can obtain photometric redshifts which can help reduce
the scatter in measuring the size variance induced by lensing,  
allow one to check for intrinsic correlations in sizes
and eliminate their contribution if needed. 
It also allows for the possibility of measuring the evolution of matter
clustering by binning the source galaxies in redshift (Jain \& Seljak 1997;
Hu 1999). With a psf of order 0.1 arcseconds and deep imaging, 
it is feasible to make size measurements
on of order a million galaxies over a 10 square degree area (based on 
the size vs. magnitude measurements in the Hubble Deep Field by 
Gardner \& Satyapal 2000). This would
give adequate ${\cal S/N}$ to measure the variance of the size 
distribution over a few bins in angle ranging from 1 to 10 arcminutes. 

With an area coverage of 100s of square degrees, 
which would probably be feasible only with a
dedicated imaging satellite such as SNAP, one can measure the projected
matter power spectrum to a precision of a few percent, measure higher
order correlations, and ideally in combination with shear information, 
get useful constraints on cosmological parameters. 
On the smallest scales, cross-correlation
statistics would probe galaxy halos on scales of a few 10s of Kpc. By
combining the magnification measurements with the shear, the density profile of
halos can be measured far more accurately than with just galaxy-galaxy
lensing, which probes only the integrated mass within radii. Further work is
needed to quantify this, explore how small the scales that one can
probe are, and check the validity of the approximation of equation 1
on these scales. Magnification effects make possible other useful
measures of the non-Gaussian lensing field that have proven difficult to
obtain from shear data, such as the skewness of $\kappa$ which probes 
$\Omega_m$ (Bernardeau et al 1997) and
peak statistics which probe the mass function of halos (Jain \& van
Waerbeke 2000). 

The cross-correlation effects of magnification on the number densities
of galaxies, and of foreground galaxy position with background galaxy
sizes, are in principle easier to measure. This is because these 
statistics are first order in the lensing convergence whereas the size 
variance is of second order. 
The interpretation is more complex in that
it involves the relation of a foreground galaxy population with the mass. 
The main requirements
for accurate measurements are photometric redshifts for a large sample 
of galaxies (to separate the foreground and background galaxies), and 
high imaging quality as discussed above. For deep imaging data that 
has a redshift distribution peaked at $z\gsim 1$, 
an adequate dataset would encompass 10 square degrees, while an ideal 
dataset would cover more than a 100 square degrees. The southern strip 
of SDSS fulfills the requirements outlined above, as do other smaller 
imaging surveys that are in progress or being planned. 

It is hoped that the discussion and results presented here motivate
the integration of magnification measurements as part of the scientific
agenda of wide area imaging surveys. The precise requirements for a given
survey that will enable useful magnification measurements to be made need
careful consideration. At the same time work is needed on survey strategy, 
techniques for combining magnification and shear information, and 
appropriate statistical measures that can be extracted from the data. 

Acknowledgments: I would like to thank Matthias Bartelmann, 
Andrew Connolly, Gary Bernstein, Jason Rhodes, David Rusin, 
Alex Szalay, Masahiro Takada and Andy Taylor 
for helpful discussions. Comments from an anonymous refereee
led to improvements in the paper. This work was supported in part 
by a NASA-ATP grant and a Keck foundation grant.

\begin{figure}[t]
\vspace{7.5cm}
\caption{(a) The left panel shows the variance in fluctuations 
in the sizes of galaxy images vs. angle $\theta$ for a 100 square
degree survey. The solid curves 
shows the predicted variance induced by lensing 
for a $\Lambda$CDM model for mean source redshift $z_s=1$ (lower
curve) and $z_s=2$ (upper curve). 
The dashed curves shows the contribution due to intrinsic size 
dispersion for galaxy number density $n_g=40$ per square
arcminute (lower) and $n_g=20$ (upper). The dotted curve shows
the sample variance contribution for $z_s=1$ (lower) and
$z_s=2$ (upper). \ \ \ \ \ \ \ (b) The right panel shows 
the ${\cal S/N}$ expected for a 10 square degree survey (lower
solid curve), a 100 square degree survey (thick solid curve), and a 1000
square degree survey (upper solid curve) with $n_g=40$ and $z_s=1$. 
The dashed and dot-dashed 
curves show the decrease in ${\cal S/N}$ for the middle curve if only half the number 
density of galaxies is available, and the increase if $z_s=2$, respectively. 
Sparse sampling would enhance the ${\cal S/N}$ for large $\theta$ compared to the
results shown here which assume a filled survey. 
}
\includegraphics{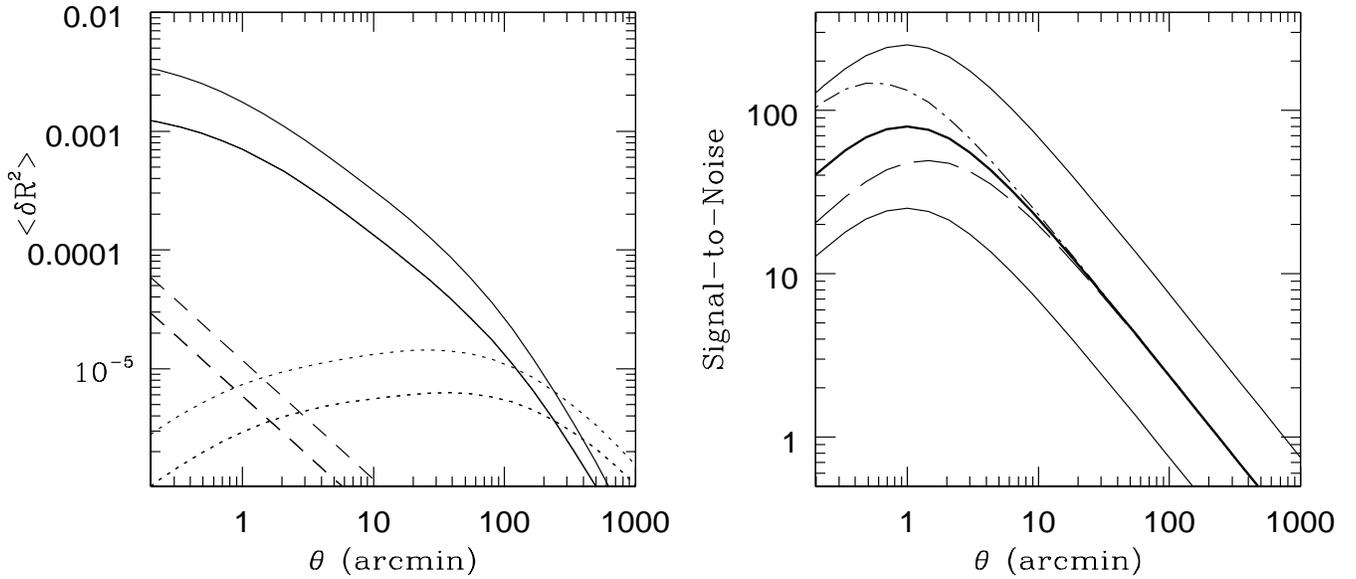}
\label{figrms}
\end{figure}

\end{document}